\documentclass[%
reprint,
showkeys,
superscriptaddress,
preprintnumbers,
nofootinbib,
amsmath,amssymb,
aps,
prd,
floatfix,
]{revtex4-2}

\usepackage{afterpage}
\usepackage{subfigure}
\usepackage{rotating} 
\usepackage{booktabs}
\usepackage{graphicx}
\usepackage{dcolumn}
\usepackage{bm}
\usepackage{float}
\usepackage{amsmath}
\usepackage{natbib}
\usepackage{hyperref}

\usepackage{xcolor}

\begin{document}

\preprint{H.~Song \& B.-Q.~Ma, \href{https://doi.org/10.1016/j.physletb.2025.139959}{Phys. Lett. B 870 (2025) 139959}}

\title{Carpet-3 300 TeV Photon Event as an Evidence for Lorentz Violation}

\author{Hanlin Song}

\affiliation{School of Physics, Peking University, Beijing 100871, China}

\author{Bo-Qiang Ma}\email{mabq@pku.edu.cn}
\thanks{corresponding author.}

\affiliation{School of Physics, Zhengzhou University, Zhengzhou 450001, China}

\affiliation{School of Physics, Peking University, Beijing 100871, China}

\begin{abstract}
The detection by the Carpet-3 Group of a 300 TeV photon, observed 4536 seconds after the prompt emission of the historic gamma-ray burst GRB 221009A, provides unprecedented opportunities to test Lorentz invariance violation (LV) at energy scales approaching the Planck regime. By analyzing the temporal and spatial properties of this ultra-high-energy photon in conjunction with lower-energy photons from other bursts and the same burst, we demonstrate consistency with subluminal LV scenarios characterized by an energy scale \( E_{\rm LV} \sim 3 \times 10^{17} \, \rm{GeV} \). This work bridges multi-year LV studies using GeV-TeV photons and establishes GRB 221009A as a pivotal laboratory for quantum spacetime phenomenology.  
\end{abstract}

\maketitle
\section{Introduction}

Gamma-ray bursts (GRBs) are astrophysical phenomena characterized by intense emissions of gamma rays, typically linked to extreme cosmic occurrences. Investigating GRBs at multi-TeV energies enables the exploration of the high-energy processes and properties of these events, providing valuable insights into the fundamental physical principles.
GRB 221009A, the so-called ``Brightest Of All Time" (BOAT) event, released unprecedented high-energy radiation, including a 12.2 TeV photon detected by LHAASO \cite{LHAASO, LHAASO:2023lkv} and a candidate 300 TeV photon reported by the Carpet-3 Group \cite{Carpet-3Group:2025fcs}. Conventional Extragalactic Background Light (EBL) absorption models predict the attenuation of photons above 10 TeV at the redshift of this event (\(z \sim 0.151\)) \cite{Li:2022vgq, Li:2022wxc}. However, both detections deviate from these predictions. 
Lorentz invariance violation (LV) - a key feature of quantum gravity theories - offers a compelling framework to resolve this discrepancy  \cite{Li:2023rgc,Li:2023rhj}. LV modifies the kinematics of photon propagation \cite{Li:2021cdz}, potentially suppressing \(\gamma\gamma \to e^+e^-\) interactions at extreme energies.

The Carpet-3 Group reported a significant observation of a 300 TeV photon-induced air shower event associated with GRB 221009A \cite{Carpet2, Carpet-3Group:2025fcs}. This represents the highest-energy photon event ever associated with a GRB, providing crucial insights into high-energy astrophysics and potential new physics. The event shows strong spatial alignment with GRB 221009A, within an angular resolution of less than 1.8°. The probability of such a photon-like event coinciding with a background photon is approximately 0.9\% (\(\sim9 \cdot 10^{-3}\)), confirming a significant connection with GRB 221009A. Using upgraded detectors and machine learning analysis, the initial energy of the photon is reconstructed as \(300^{+43}_{-38}\) TeV, updating previous preliminary records (such as the 251 TeV event reported by Carpet-2 \cite{Carpet2}).

The Carpet-3 300 TeV photon event was detected 4536 seconds after the initial burst trigger of GRB 221009A. This substantial time lag offers a unique opportunity to investigate Lorentz invariance violation (LV) by analyzing the time delay between this ultra-high-energy photon and lower-energy photons emitted during the prompt phase. The approach of using GRB photon time delays to study LV was first proposed by Amelino-Camelia {\it et al.} \cite{Amelino-Camelia:1996bln, Amelino-Camelia:1997ieq}.

Recent studies of LV have made significant progress through multi-year analyses of GRB data. In particular, a recent work \cite{song_and_ma_ApJ} developed a unified framework for Lorentz violation signatures. This is achieved by applying an energy-dependent intrinsic time-delay model \cite{Song:2024and} to 14 multi-GeV photons from the Fermi Gamma-ray Space Telescope (FGST) and three notable photons: the 99.3 GeV photon of GRB 221009A observed by FGST \citep{Lesage:2023vvj, Zhu:2022usw}, the 1.07 TeV photon of GRB 190114C observed by the Major Atmospheric Gamma Imaging Cherenkov (MAGIC) telescope \cite{MAGIC:2019lau, MAGIC:2020egb}, and the 12.2 TeV photon of GRB 221009A observed by the Large High Altitude Air-shower Observatory (LHAASO) \cite{LHAASO:2023lkv}. This model separates astrophysical emission mechanisms from potential LV effects, allowing for the reliable extraction of LV-induced time delays \cite{song_ma_MC}.

In this study, we show that the Carpet–3 300 TeV photon event is highly consistent with the LV phenomenology observed in both the 14 FGST photons \cite{Song:2024and} and the combined 14 + 3 high-energy photon sample \cite{song_and_ma_ApJ}. The temporal and spatial characteristics are in line with a subluminal LV scenario, with an energy scale of \( E_{\rm LV} \sim 3 \times 10^{17} \, \rm{GeV} \) as derived in \cite{Song:2024and,song_and_ma_ApJ}. This energy scale implies a suppression of LV effects at energies below \(\sim 10^{17} \, \rm{GeV} \), which is consistent with the theoretical predictions of quantum spacetime foam models \cite{Amelino-Camelia:1996bln, Amelino-Camelia:1997ieq, Ellis:1999rz, Ellis:1999uh, Ellis:2008gg, Li:2009tt, Li:2021gah, Li:2021eza} and/or loop quantum gravity models \cite{Gambini:1998it, Alfaro:1999wd,Li:2022szn}.

\section{Time delay models and parameter estimation methods}

From Lorentz invariance violation scenario, for photons with energies significantly lower than the Planck energy, the dispersion relation needs to be modified in a model-independent manner, leading to the expression \cite{He:2022gyk,Xiao:2009xe},
\begin{equation}
E^2 \simeq p^2c^2\left[1-s_n\left(\frac{pc}{E_{\mathrm{LV},n}}\right)^n\right],
\end{equation}
where $p$ represents the momentum of the photon, 
$E_{{\rm{LV}},n}$ denotes the $n$-th order energy scale of Lorentz invariance violation, and $s_n \equiv \pm 1$ indicates whether superluminal ($s_n = -1$) or subluminal ($s_n = +1$). The group velocity of the photon can be expressed as,
\begin{equation}
v \simeq c\left[1-s_n\frac{n+1}{2}\left(\frac{pc}{E_{\mathrm{LV},n}}\right)^n\right].
\end{equation}

Thus, in the the framework of LV, the observed time delay between high- and low-energy photons from GRBs can be divided into two parts \cite{Ellis:2005sjy, Shao:2009bv, Zhang:2014wpb, Xu:2016zxi, Xu:2016zsa, Zhu:2021pml, Zhu:2021wtw, Zhu:2022usw, Huang:2019etr},
\begin{equation}
	\label{obsdelay}
	\Delta t_{\mathrm{obs}}=\Delta t_{\mathrm{LV}}+(1+z)\Delta t_{\mathrm{in}},
\end{equation}
where $\Delta t_{\mathrm{LV}}$ denotes the time delay due to LV, $\Delta t_{\mathrm{in}}$ is the intrinsic time delay at the source, and $z$ signifies the redshift of the corresponding GRB. Due to the expansion of the Universe, the LV time delay can be expressed as \cite{Jacob:2008bw, Zhu:2022blp}:
\begin{equation}
	\label{lorentzdelay}
	\Delta t_{\mathrm{LV}}=s_{n}\frac{1+n}{2H_{0}}\frac{E_{\mathrm{high,o}}^{n}-E_{{\rm low,o}}^{n}}{E_{\mathrm{LV},n}^{n}}\int_{0}^{z}\frac{(1+z')^{n}\mathrm{d}z'}{\sqrt{\Omega_{m}(1+z')^{3}+\Omega_{\Lambda}}},
\end{equation}
where $\Omega_{m}$ and $\Omega_{\Lambda}$ are the matter density and energy density parameters from the $\Lambda$CDM model, respectively.

For the intrinsic component, previous studies have considered a common intrinsic time delay term \cite{Shao:2009bv, Zhang:2014wpb, Xu:2016zxi, Xu:2016zsa, Liu:2018qrg,Zhu:2021pml, Zhu:2021wtw, Zhu:2022usw}. However, a recent work proposes a new intrinsic time delay model, which assumes that high-energy photons ranging from GeV to TeV are emitted at different times at source, depending on the energy value~\cite{Song:2024and}, 
\begin{equation}
    \Delta t_{\rm{in}} = \Delta t_{\rm in, c} + \alpha E_{\rm s},
\end{equation}
where $\alpha$ is the coefficient and $E_{\rm s}$ is source frame energy. 
Following previous works~\citep{Xu:2016zxi, Xu:2016zsa}, we adopt $n = 1$ and $s_n = +1$, so that Eq.~(\ref{obsdelay}) can be rewritten as,
\begin{equation}
\label{sourcedelay}
\frac{\Delta t_{\mathrm{obs}}}{1+z} = \Delta t_{\mathrm{in}} + a_{\rm LV}K_1,
\end{equation}
where $a_{\rm LV} = 1/E_{\rm LV}$ and $K_1$ is
\begin{equation}
K_1=\frac{1}{H_0}\frac{E_{\rm high,o} - E_{\rm low,o}}{1+z}\int_0^z\frac{(1+z')\mathrm{d}z'}{\sqrt{\Omega_\mathrm{m}(1+z')^3+\Omega_\Lambda}}.
\end{equation}
Refs.~\cite{Song:2024and, song_and_ma_ApJ} applied this model to the analysis of 14 multi-GeV photons, which have been studied in previous works~\cite{Xu:2016zxi, Xu:2016zsa}, as well as three remarkable photons with energy of 99.3 GeV, 1.07 TeV, and 12.2 TeV. Consistent results are obtained with the new intrinsic time delay models in two aspects. On the one hand, both the previous and the new models yield a consistent value of $E_{\rm LV} \simeq 3 \times 10^{17}$ GeV for 14 multi-GeV photons~\cite{Song:2024and}. On the other hand, consistent results are also obtained~\cite{song_and_ma_ApJ} for all parameters when analyzing the 17 photons dataset (including TeV and GeV photons) and the 14 photons dataset (only GeV photons) with the new intrinsic time delay model. Moreover, using the the Akaike information criterion (AIC) criterion \cite{akaike1981likelihood, Biesiada:2009zz}, the new model demonstrates better performance in fitting both datasets than the previous model \cite{Song25Dark}.  The new model is also tested through Monte Carlo simulation of GRB data as a reliable framework for future searches related to Lorentz violation \cite{song_ma_MC}.

We adopt the same Bayesian framework for parameter estimation under a Gaussian noise model as used in Refs.~\cite{Song:2024and, song_and_ma_ApJ, Song25Dark}. The posterior can be expressed as, 
\begin{equation}
\begin{aligned}
     p \propto & \prod_{j=1} \frac{1}{\sqrt{2\pi\left(a_{\rm LV}^2\sigma_{K_{1,j}}^2 + \alpha^2\sigma_{E_{{\rm high,s},j}}^2 + \upsilon^2 + \sigma_{y_{j}}^2 \right)}}\\
     & \times \exp{\left(-
 \frac{\left(\frac{\Delta t_{\mathrm{obs},j}}{1+z_{j}} - a_{\rm LV}{K_{1,j}} - \alpha E_{{\rm high,s},j} - \mu\right)^2}{2\left(a_{\rm LV}^2\sigma_{K_{1,j}}^2 + \alpha^2\sigma_{E_{{\rm high,s},j}}^2 + \upsilon^2 + \sigma_{y_{j}}^2 \right)}\right)} \\
& \times p\left(a_{\rm LV}\right)p(\alpha)p\left(\mu\right)p\left(\upsilon\right), 
\end{aligned}
\end{equation}
where $\Delta t_{\rm in,c}$ is assumed to follow a Gaussian distribution $p\left(\Delta t_{\rm in,c} \right) \sim \mathcal{N} \left(\mu, \upsilon^2\right)$, in which $\mu$ represents the mean value of the common intrinsic time delay, allowing the emission to occur over a finite time interval (denoted by $\upsilon$). The priors $p\left(a_{\rm LV}\right), \ p(\alpha), \ p\left(\mu\right), \ p\left(\upsilon\right)$ represent the prior of the four parameters and are assumed to follow flat distributions, 
\begin{align}
    \begin{cases}
        p\left(a_{\rm LV}\right) \sim U\left[0, 30 \right] \times 10^{-18} \ \rm{GeV^{-1}}, \\
        p(\alpha) \sim U \left[-30, 30 \right] \ {\rm s \cdot GeV^{-1}}, \\
        p\left(\mu\right) \sim U\left[-30, 30 \right] \ {\rm s}, \\
        p\left(\upsilon\right) \sim U\left[0, 30 \right] \ {\rm s}. \\
    \end{cases}
    \label{priors}
\end{align}

We employ the \texttt{bilby} package \citep{Ashton:2018jfp, Romero-Shaw:2020owr} to perform the parameter estimation.

\section{Results}

In this study, we examine the newly detected 300 TeV Carpet-3 photon under two scenarios. First, we combine it with a dataset of 14 multi-GeV photons to test its consistency with the GeV-band data. Subsequently, we analyze it in conjunction with a 17-photon dataset to assess its compatibility with both GeV- and TeV-band photons. The new Carpet-3 photon arrived 4536 seconds after the Fermi-GBM trigger of GRB 221009A and has an energy resolution of 13\% \cite{Carpet-3Group:2025fcs}. The details of the 14-photon and 17-photon datasets can be found in Ref.~\cite{Song:2024and}. The results are presented in Table~\ref{results_table}.

\begin{table*}[ht]
  \centering
  \caption{Table of estimated parameters for Fig.~\ref{14_photons} and Fig.~\ref{17_photons}, where the error bars represent the 1$\sigma$ uncertainties. }
    \begin{tabular}{cccccc}
    \toprule
    Case & $a_{\rm LV} ~ (10^{-18} ~ {\rm GeV}^{-1})$ & $\alpha ~( \rm{s} \cdot {\rm GeV}^{-1}) $ & $\mu ~({\rm s})$ & $\upsilon ~({\rm s})$ & $E_{\rm LV} ~(10^{17}~ {\rm GeV})$ \\
    \midrule
    14 photons & $3.28^{+1.08}_{-0.95}$ & $-0.15^{+0.10}_{-0.11}$ & $-4.53^{+4.27}_{-4.43}$ & $5.30^{+1.97}_{-1.47}$ & $3.04^{+1.23}_{-0.76}$ \\
    14 + Carpet-3 & $3.31^{+0.93}_{-0.81}$ & $-0.16^{+0.05}_{-0.07}$  & $-3.39^{+3.42}_{-3.61}$ & $5.03^{+1.81}_{-1.39}$ & $3.02^{+0.98}_{-0.66}$ \\
    17 photons & $3.33^{+0.91}_{-0.91}$ & $-0.20^{+0.07}_{-0.07}$ & $-1.04^{+3.55}_{-3.47}$ & $5.10^{+2.04}_{-1.51}$ & $2.99^{+1.11}_{-0.65}$  \\
    17 + Carpet-3 & $3.02^{+0.91}_{-1.07}$ & $-0.17^{+0.07}_{-0.06}$ & $-1.43^{+3.54}_{-3.28}$ & $5.26^{+2.14}_{-1.53}$ & $3.30^{+1.80}_{-0.77}$  \\
    \bottomrule
    \end{tabular}%
  \label{results_table}%
\end{table*}

\begin{figure*}[ht] 
	\centering 
    \begin{minipage}{0.46\textwidth}
        \centering
        \includegraphics[width=0.95\linewidth]{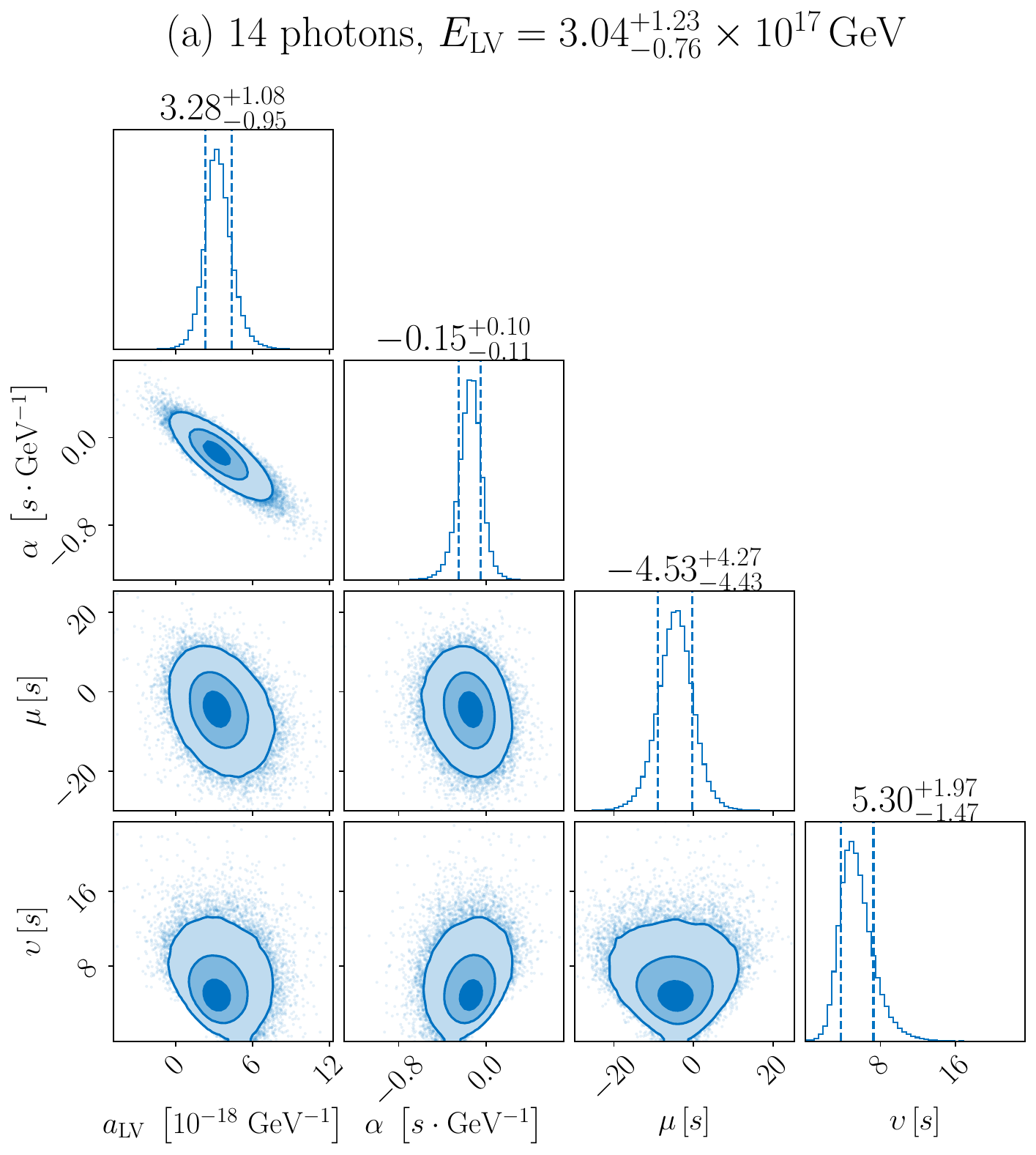}
    \end{minipage}\hfill
    \begin{minipage}{0.49\textwidth}
        \centering
        \includegraphics[width=0.95\linewidth]{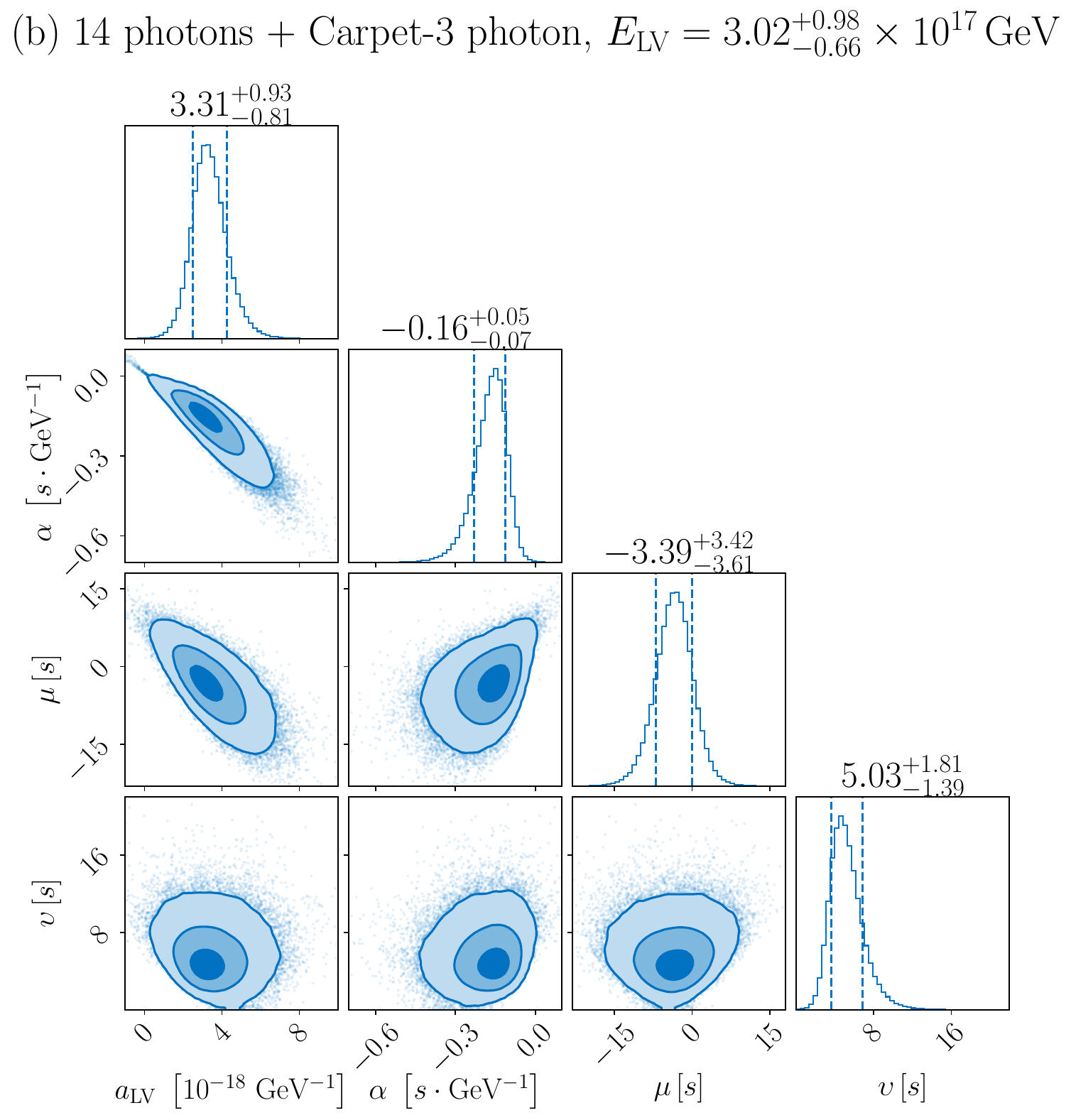} 
    \end{minipage}
    \caption{Left panel shows the re-run results for analyzing 14 FGST photons in Ref.~\cite{Song:2024and}, while the left panel shows the results for analyzing 14 FGST photons together with the new Carpet-3 photon. The 2D contours represent with different credible intervals, denoting the 1$\sigma$, 2$\sigma$, and 3$\sigma$ regions, while the vertical lines indicate the  1$\sigma$ region for the 1D marginalized posterior distribution.} 
    \label{14_photons}
\end{figure*}

The results of analyzing the new 300 TeV Carpet-3 photon in combination with the 14 FGST photons are presented in Fig.~\ref{14_photons}. The left-hand panel displays the re-run results for the 14 multi-GeV photons dataset \cite{song_and_ma_ApJ}, which provide the following parameter estimates: the LV parameter $a_{\rm LV}=3.28^{+1.08}_{-0.95}\times 10^{-18}\ \rm{GeV^{-1}}$, the energy-dependent parameter $\alpha=-0.15^{+0.10}_{-0.11}\ {\rm s \cdot GeV^{-1}}$, the mean value of the common intrinsic time delay $\mu=-4.53^{+4.27}_{-4.43}$ s, the standard deviation of the common intrinsic time delay $\upsilon = 5.30^{+1.97}_{-1.47}$ s, and the corresponding LV energy scale $E_{\rm LV}=1/a_{\rm LV}=3.04^{+1.23}_{-0.76}\times 10^{17}\ {\rm GeV}$. The right-hand panel shows the results for the joint analysis of the 14 multi-GeV photons and the new Carpet-3 photon. The estimated parameters are: the LV parameter $a_{\rm LV}=3.31^{+0.93}_{-0.81}\times 10^{-18}\ \rm{GeV^{-1}}$, the energy-dependent parameter $\alpha=-0.16^{+0.05}_{-0.07}\ {\rm s \cdot GeV^{-1}}$, the mean value of the common intrinsic time delay $\mu=-3.39^{+3.42}_{-3.61}$ s, the standard deviation of the common intrinsic time delay $\upsilon = 5.03^{+1.81}_{-1.39}$ s, and the corresponding LV energy scale $E_{\rm LV}=3.02^{+0.98}_{-0.66}\times 10^{17}\ {\rm GeV}$. All four parameters are consistent within 1$\sigma$ between the two scenarios, indicating that the new Carpet-3 photon is compatible with the previous 14 multi-GeV photons. Moreover, both scenarios suggest an LV energy scale of approximately \( \sim 3\times 10^{17} \, \rm{GeV} \).

\begin{figure*}[ht] 
	\centering 
    \begin{minipage}{0.46\textwidth}
        \centering
        \includegraphics[width=0.95\linewidth]{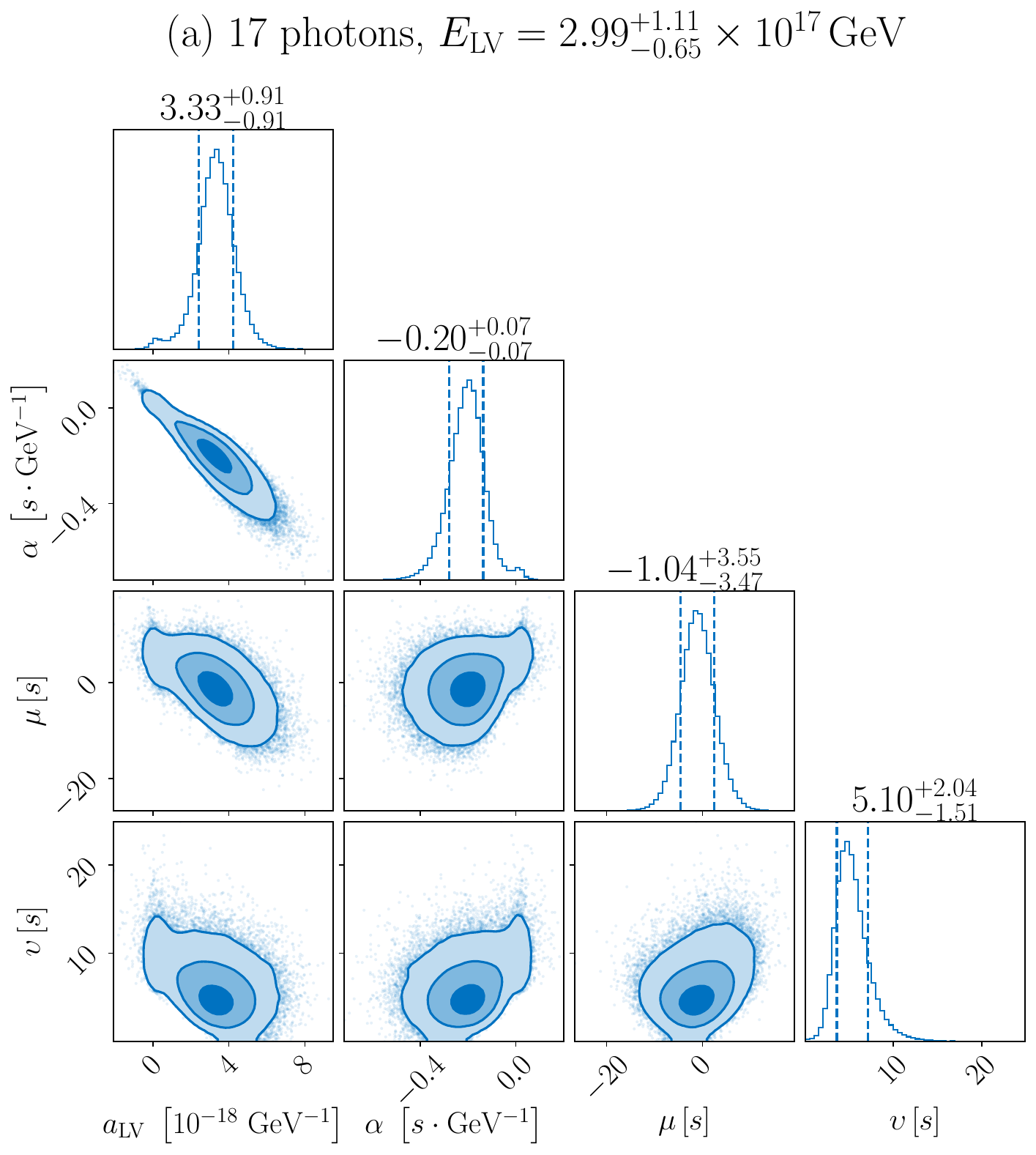}
    \end{minipage}\hfill
    \begin{minipage}{0.49\textwidth}
        \centering
        \includegraphics[width=0.95\linewidth]{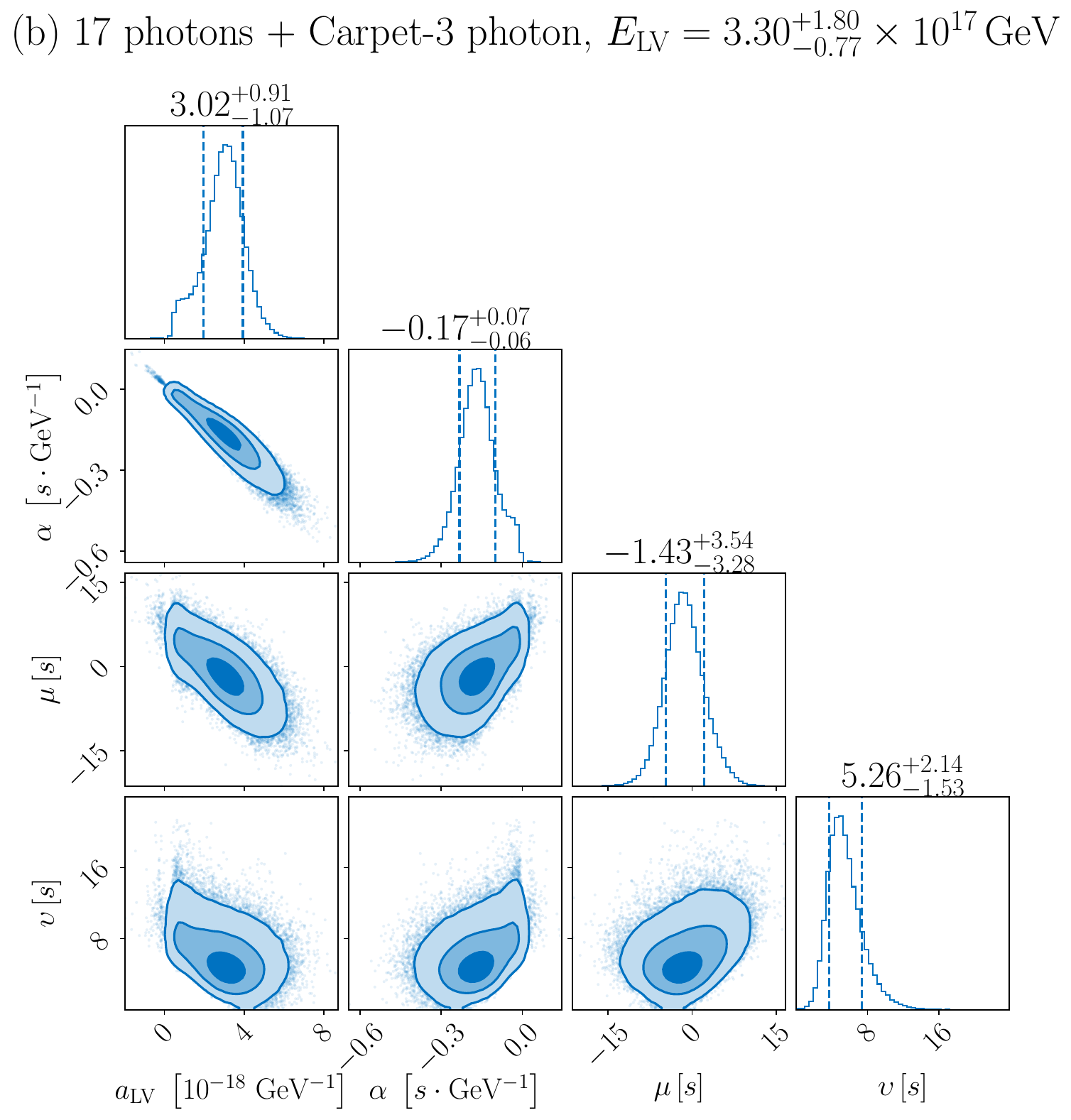} 
    \end{minipage}
    \caption{Same as Fig.~\ref{14_photons}, but for the 14 FGST photons and the three remarkable photons \cite{song_and_ma_ApJ}, including the 99.3 GeV photon of GRB 221009A observed by FGST 
\citep{Lesage:2023vvj, Zhu:2022usw}, the 1.07 TeV photon of GRB 190114C observed by the MAGIC \cite{MAGIC:2019lau, MAGIC:2020egb}, and the 12.2 TeV photon of GRB 221009A observed by the LHAASO \cite{LHAASO:2023lkv}.} 
    \label{17_photons}
\end{figure*}

The results of analyzing the new 300 TeV Carpet-3 photon along with the 17-photon dataset are shown in Fig.~\ref{17_photons}. The left-hand panel shows the re-run results for the 17-photon dataset \cite{song_and_ma_ApJ}, yielding the following estimates: the LV parameter $a_{\rm LV}=3.33^{+0.91}_{-0.91}\times 10^{-18}\ \rm{GeV^{-1}}$, the energy-dependent parameter $\alpha=-0.20^{+0.07}_{-0.07}\ {\rm s \cdot GeV^{-1}}$, the mean value of the common intrinsic time delay $\mu=-1.04^{+3.55}_{-3.47}$ s, the standard deviation of the common intrinsic time delay $\upsilon = 5.10^{+2.04}_{-1.51}$ s, and the corresponding LV energy scale $E_{\rm LV}=2.99^{+1.11}_{-0.65}\times 10^{17}\ {\rm GeV}$. The right-hand panel shows the results for the analysis of both the 17-photon dataset and the new Carpet-3 photon. The inferred parameters are: $a_{\rm LV}=3.02^{+0.91}_{-1.07}\times 10^{-18}\ \rm{GeV^{-1}}$, the energy-dependent parameter $\alpha=-0.17^{+0.07}_{-0.06}\ {\rm s \cdot GeV^{-1}}$, the mean value of the common intrinsic time delay $\mu=-1.43^{+3.54}_{-3.28}$ s, the standard deviation of the common intrinsic time delay $\upsilon = 5.26^{+2.14}_{-1.53}$ s, and the corresponding LV energy scale $E_{\rm LV}=3.30^{+1.80}_{-0.77}\times 10^{17}\ {\rm GeV}$. The two scenarios are once again consistent with each other, suggesting that the new Carpet-3 photon is compatible with both GeV and TeV photons, including the 99.3 GeV photon and 12.2 TeV photon from the same GRB 221009A. Moreover, the LV energy scale of approximately \( \sim 3\times 10^{17} \, \rm{GeV} \) is confirmed once more.

All the inferred parameters from the above four scenarios are self-consistent with each other within \(1\sigma\). This high degree of self-consistency demonstrates the robustness of the new time delay model. The model is capable of accurately explaining both the intrinsic emission mechanism of high-energy photons and the LV energy scale.

Furthermore, the consistency between the new Carpet-3 photon and the 17 other photons ranging from GeV to TeV, including the 99.3 GeV and 12.2 TeV photons from GRB 221009A, provides strong support for a significant correlation between the new Carpet-3 photon and GRB 221009A. This correlation is a crucial finding as it helps to establish the origin of the new photon and enhances our understanding of the high-energy emission processes associated with GRBs.

\section{Conclusion}
The article focuses on the detection of a 300 TeV photon-like air shower event by the Carpet-3 experiment coincident with GRB 221009A, which challenges conventional extragalactic photon propagation models. Here is the summary:

1. {\bf{Observation and motivation}}: Carpet-3 detected a 300 TeV photon 4536 seconds after the prompt emission of GRB 221009A~\cite{Carpet-3Group:2025fcs}. Traditional cosmic background light (EBL) absorption models predict attenuation of photons above 10 TeV at the redshift of GRB 221009A (\(z \sim 0.151\))~\cite{Li:2022vgq, Li:2022wxc}, yet this and other high-energy photon detections defy these expectations. Lorentz invariance violation (LV) is proposed as a mechanism~\cite{Li:2021cdz} to explain the survival of such ultra-high-energy (UHE) photons against EBL absorption~\cite{Li:2023rgc,Li:2023rhj}.

2. {\bf{LV framework and analysis method}}: LV modifies photon propagation kinematics. The observed time delay between high- and low-energy photons from GRBs in the LV framework is divided into two parts: the time delay due to LV (\(\Delta t_{\mathrm{LV}}\)) and the intrinsic time delay at the source \((1 + z)\Delta t_{\mathrm{in}}\). A new intrinsic time delay model \(\Delta t_{\rm{in}} = \Delta t_{\rm in, c} + \alpha E_{\rm s}\) is considered~\cite{Song:2024and}, which assumes high-energy photons from GeV to TeV are emitted at different times depending on energy. A Bayesian framework for parameter estimation under a Gaussian noise model is adopted, and the \texttt{bilby} package is used for parameter estimation.

3. {\bf{Results}}: Analyzing the 300 TeV Carpet-3 photon with two scenarios (14 multi-GeV photons dataset~\cite{Song:2024and} and 17 photons dataset including GeV and TeV photons~\cite{song_and_ma_ApJ}). In both scenarios, the estimated parameters (LV parameter \(a_{\rm LV}\), energy-dependent parameter \(\alpha\), mean value of common intrinsic time delay \(\mu\), standard deviation of the common intrinsic time delay \(\upsilon\), and LV energy scale \(E_{\rm LV}\)) are consistent within 1$\sigma$. The LV energy scale is found to be around \( \sim 3 \times 10^{17} \, \rm{GeV} \) in all cases. This indicates that the new Carpet-3 photon is compatible with previous multi-GeV photons~\cite{Song:2024and} and both GeV and TeV photons from GRB 221009A and other bursts~\cite{song_and_ma_ApJ}, and supports a strong correlation between the new Carpet-3 photon and GRB 221009A. The new time-delay model is robust in explaining both the intrinsic emission mechanism of high-energy photons and the LV energy scale. 

Overall, this work bridges multi-messenger astrophysics and quantum gravity phenomenology, offering a novel way to probe Planck-scale physics and establishing GRB 221009A as an important laboratory for quantum spacetime phenomenology. 
\\

\emph{\textbf{Acknowledgements}.---}%
This work is supported by National Natural Science Foundation of China under Grant No.~12335006. This work is also supported by High-performance Computing Platform of Peking University.

\bibliographystyle{elsarticle-num}
\bibliography{scibib}
\end{document}